%% file: sb_paper.tex
\documentstyle[fleqn]{article}     
\input def                         

\def\ltsim{$\mathrel{\spose{\lower 3pt\hbox{$\sim$}}
        \raise 2.0pt\hbox{$<$}}$}
\def\gtsim{$\mathrel{\spose{\lower 3pt\hbox{$\sim$}}
        \raise 2.0pt\hbox{$>$}}$}

\def\ApJ{{\it Astrophys.~J.}}

\def\ApJS{{\it Astrophys.~J.~Suppl.}}

\def\MN{{\it Mon.~Not.~R.~Astron.~Soc.}}

\def\PASJ{{\it Publ.~astr.~Soc.~Japan}}

\def\Sc{{\it Science}}

\begin{document}                            

\vspace*{-10mm}
\vspace*{10mm}
\begin{center}
{\LARGE The Imprint of Galaxy Formation on X-ray Clusters}\\[10mm]
{\large Trevor J. Ponman$^*$, Damian B. Cannon$^*$ 
\& Julio F. Navarro$^\dag$ }\\[5mm]
{$^*$School of Physics \& Astronomy, University of Birmingham,\\Edgbaston,
Birmingham B15 2TT, UK\\
$^\dag$ Department of Physics \& Astronomy, University of Victoria,\\
Victoria, BC V8P 1A1, Canada}
\end{center}
{\bf It is widely believed that structure in the Universe evolves
hierarchically, as primordial density fluctuations, amplified by gravity,
collapse and merge to form progressively larger systems. The structure and
evolution of X-ray clusters, however, seems at odds with this hierarchical
scenario for structure formation$^{1}$. Poor clusters and groups, as well as
most distant clusters detected to date, are substantially fainter than expected
from the tight relations between luminosity, temperature and redshift predicted
by these models. Here we show that these discrepancies arise because, near the
centre, the entropy of the hot, diffuse intracluster medium (ICM) is higher than
achievable through gravitational collapse, indicating substantial
non-gravitational heating of the ICM.  We estimate this excess entropy for the
first time, and argue that it represents a relic of the energetic winds through
which forming galaxies polluted the ICM with metals. Energetically, this is only
possible if the ICM is heated at modest redshift ($z $\ltsim $2$) but prior to
cluster collapse, indicating that the formation of galaxies precedes that of
clusters and that most clusters have been assembled very recently.}\\[2mm]

Recent numerical work has unveiled a remarkable similarity in the structure of
galaxy clusters formed through hierarchical clustering: properly scaled, the
dark matter density profiles of virialized clusters are nearly
identical$^{2}$. The scaling prescription is simple. Coeval clusters have
approximately the same characteristic density$^{3}$, $M/r^3=$ constant, where
$M$ and $r$ are the characteristic mass and radius of a cluster.  The mass
profiles of all clusters are then similar once radii are scaled to a ``virial''
radius, $r_v \propto (GM/r)^{1/2}$. Baryons in galaxy clusters contribute a
minority of the total mass, and are found in a hot, diffuse, X-ray bright
intracluster medium at a temperature, $T$, which indicates the depth of the
potential well; $T \propto GM/r \propto r_v^2$. If gas traces mass, scaled X-ray
surface brightness profiles are also expected to be similar. Furthermore, a
simple scaling is expected between X-ray luminosity, $L_X$, and
temperature. Assuming that the emission is dominated by bremsstrahlung, $L_X
\propto M_{\rm gas}^2 \, R^{-3} \, T^{1/2}$, or $L_X \propto f_{gas}^2 \,
T^{2}$, where $f_{gas}=M_{gas}/M$ is the gas mass fraction.

X-ray clusters deviate substantially from this simple scaling, and the observed
$L_X$:$T$ relation$^{4}$ is considerably steeper than $T^2$.  Two main
interpretations are usually considered: either the gas fraction is
systematically higher in hotter clusters$^{5}$, or the spatial distribution of
the gas deviates systematically from the similarity profile in clusters of
different temperature. A simple example of the latter is provided by the
``preheated ICM'' model$^{1,6}$, wherein the entropy of the gas in all clusters
is raised at early times to levels comparable to those reached through
gravitational collapse. This entropy ``floor'' imposes a maximum central density
for the ICM which increases roughly as $T^{3/2}$ and therefore affects cooler
clusters more severely, reducing their X-ray luminosities and steepening the
$L_X$:$T$ relation relative to the similarity scaling.

In order to investigate the nature of this similarity breaking, we have
extracted X-ray surface brightness profiles from ROSAT observations of 25
systems spanning a factor $\sim 15$ in temperature, from rich clusters to
small groups. The sample is restricted to systems with reasonably regular
X-ray morphology and good estimates of X-ray temperature from the GINGA
or ROSAT satellites. The profiles are then scaled so that they would
coincide if obeying similarity, and are shown in Figure 1. (Details of
the scaling procedure are given in the caption.) Profiles are
colour-coded according to system temperature. Systems with $T>4$\keV\
(shown in blue) have on average similar profiles and show no strong trend
with temperature, whilst cooler systems (green and red) have profiles
that become progressively shallower and less centrally concentrated. This
is especially noticeable outside the centre, where cooling flows dominate
and the profiles of most systems steepen ($r \, $\ltsim$ \, 0.05 r_v$).
Deviations in the scaled profiles persist out to radii where cooling
timescales greatly exceed the age of the universe, so it seems unlikely
that radiative effects are responsible for this result.

The trends in Figure 1 are clearly inconsistent with simple variations in
overall baryon fraction (which would result in profiles similar in shape, but
shifted vertically with $T$), but are in good agreement with the predictions of
the ``preheated ICM'' model described above. However, one can imagine other
effects which would lead to a reduction in central gas densities. For example,
non-thermal pressure support (e.g. by magnetic fields) or a failure in the
similarity of the underlying gravitational potentials. The unmistakable
signature of preheating is a rise in the entropy of the gas. Magnetic support,
for example, would actually reduce the entropy, since it would weaken the shocks
which heat the gas as the cluster potential develops.  To explore this, we have
fitted simple isothermal models to the X-ray emission profiles in order to
derive profiles of the ``entropy'', $S(r)=T/n_e^{2/3}$, for all systems. The
fits exclude any central excess arising from a cooling flow, and employ the
widely-used isothermal $\beta$-model$^{7}$ to derive $n_e(r)=n_e^0 [1+
(r/r_{core})^2]^{-3\beta/2}$.

Figure 2 shows the entropy of the gas at a fiducial radius of $0.1 \, r_v$, just
outside the cooling flow-dominated region. If the gas profiles were similar, the
entropy at a fixed fraction of $r_v$ would be directly proportional to the
temperature (solid line).  Cool ($T<4$ keV) clusters clearly have entropies
higher than achievable through gravitational collapse alone.  The presence of
large cores in the entropy profiles of a small sample of galaxy groups and
clusters has been noted previously$^{8}$. Here we see that this is part of a
systematic trend whereby low mass systems progressively depart from the
self-similar scaling, and that at $T \sim 1$ keV the entropy appears to converge
to a characteristic floor value of $\sim 100 \, h^{-1/3}$ keV cm$^2$.

What processes are potentially able to establish this trend?  Cooling flows will
remove low-entropy gas from the centre allowing higher entropy material to flow
inwards. However, an effect of the magnitude seen can only be obtained if the
amount of gas removed by the cooling flow is a substantial fraction of the ICM,
which seems unlikely$^{9}$. Moreover, substantial loss of gas to cooling flows,
or large systematic variations in galaxy formation efficiency, are inconsistent
with the fact that the gas fraction within $r_v$, and the metallicity of the
ICM, are roughly independent of the temperature of the system$^{10,11}$.  It is
clear from the observed metal content of the ICM that galaxies have polluted
their surroundings with substantial quantities of metal-rich gas. The strong
galactic winds responsible for this pollution are thus the most natural
mechanism available to heat the ICM. In hot clusters the gravitational collapse
of the system generates entropies in excess of the floor value established by
galaxy winds, but in cooler systems it is preserved during collapse, and
prevents the gas from collapsing to high central densities$^{12,13}$.

The energy requirements of the preheating mechanism can be assessed by
considering the Coma cluster, the best studied nearby rich cluster, although the
argument applies equally to smaller systems. The total stellar mass in galaxies
in Coma is $\sim 10^{13}
\, h^{-1} M_{\odot}$ and therefore the total energy available from
supernovae is $\sim 10^{62} \, h^{-1}$ erg (assuming 1 SN event per $100 \,
M_{\odot}$ of stars formed). Since the gas mass in the ICM is $ \sim 5.5 \times
10^{13} \, h^{-5/2} M_{\odot}$, these supernovae can raise the ICM temperature
by $\sim 0.4 \, h^{3/2}$ keV. Hence SN can only raise the entropy, $T
n_e^{-2/3}$, to the observed floor value of $\sim 100\, h^{-1/3} \, {\rm keV} \,
{\rm cm}^2$ if $n_e < 2.5 \times 10^{-4} \, h^{11/4}$ cm$^{-3}$, a value well
below the electron density at our fiducial radius of $r=0.1r_v$ in
Coma$^{14}$. In other words, dumping all of the available supernova energy into
the ICM of Coma {\it today} would be insufficient to heat the denser inner
regions to the desired entropy floor. The problem is even more acute at higher
redshift, since the density of the Universe, and that of all collapsed clumps,
is higher then.  This simple analysis implies that the gas in the ICM cannot be
in collapsed, overdense systems at the time of preheating. The formation of
galaxies must therefore precede the collapse of the cluster as a whole,
validating the basic tenet of the hierarchical clustering paradigm.

When did this heating occur?  Assuming that the gas is uniformly
distributed throughout the Universe at the time of preheating, the
density limit gives a strict upper limit on the preheating redshift,
$z_{\rm preh} \, $\ltsim$ \, 10$. In practice, density variations and
reductions in heating efficiency reduce this redshift limit. An
entropy of $100 h^{-1/3}$ keV cm$^2$ is approximately the value
baryons would have at present if they were spread uniformly throughout
the Universe at the mean density inferred from models of cosmic
nucleosynthesis$^{15}$ and heated to a temperature of $30,000$ K.  The
entropy of the Ly-$\alpha$ forest gas observed in QSO spectra (the
dominant baryonic component of the Universe at $z\sim 2$)$^{16}$, is
almost an order of magnitude smaller, since the temperatures are of
the same order but the density of the Universe at $z\sim 2$ is a
factor of $\sim 30$ greater. If, as expected, winds from forming
galaxies affect cluster and field environments alike, observations of
the Ly-$\alpha$ forest imply that $z_{\rm preh} $\ltsim$ 2$,
consistent with the low redshift at which the global star formation
rate in the universe appears to have peaked$^{17}$.

We can also estimate the typical propagation velocity of the winds,
$v_w$, by noting that one galaxy can perturb a volume of order $(4 \pi/3)
(v_w/100 \, {\rm km \, s}^{-1})^3 \, h^{-3}$ Mpc$^3$ in a Hubble time.
Since the mass in the Coma cluster has been collected from a volume a few
hundred times this, whilst its optical luminosity is approximately that
of 200 galaxies like the Milky Way, most of the ICM in Coma could have
been preheated by winds blowing at $v_w $\gtsim$ 100$ km s$^{-1}$. Such
velocities lie comfortably within the range allowed by winds observed in
galaxies which are actively forming stars$^{18}$.

The results presented above demonstrate the dramatic influence that
forming galaxies have on their surroundings. In the case of rich
clusters, which account for $\sim$10\% of galaxies, preheating by galaxy
winds will substantially modify the evolution of the X-ray luminosity and
temperature functions$^{19}$, and will result in a luminosity-temperature
relation which is independent of redshift$^{2,6}$, in good agreement with
current compilations of X-ray data$^{20}$.  In poor clusters and groups,
the entropy floor will have more dramatic effects, and may reduce the
baryon fraction within the virial radius below the universal
value$^{21}$.

For the majority of galaxies which, like the Milky Way, are not located
in virialised groups or clusters, preheating will largely prevent gas
from collapsing into their potential wells, leaving the bulk of the
baryons in the Universe in a currently undetectable high entropy
sea$^{22}$. Inhibiting the collapse of baryons into galaxies
after preheating robs them of their main source of fresh fuel and may
explain the precipitous drop$^{17}$ in the global star formation rate of
the universe since $z \sim 1$. In this scenario, most baryons are
distributed between galaxies at temperatures $T \sim 30,000 \,
\delta^{2/3}$ K, where $\delta$ is the gas overdensity relative to the
universal average. This explains why most galaxies lack the prominent
X-ray halos observed in rich clusters. In the halos of galaxies like the
Milky Way, where the virial temperature is $\sim 0.1$ keV, gas densities
cannot exceed a few times the universal average, and would only be
detectable through absorption studies of highly ionized species of metals
such as Si in the ultraviolet spectra of background sources. Only in
``hotter'' systems, such as massive ellipticals, galaxy groups and
clusters, can baryons achieve densities high enough to shine profusely
and be easily detected.

{\bf References}\\[1mm]

1. Kaiser, N. Evolution of clusters of galaxies. \ApJ\ {\bf 383}, 
104--111 (1991).

2. Navarro J.F., Frenk C.S. \& White S.D.M. Simulations of X-ray clusters.
\MN\ {\bf 275}, 720--740 (1995).

3. Navarro J.F., Frenk C.S. \& White S.D.M. A universal density profile
from hierarchical clustering. \ApJ\ {\bf 490}, 493--508 (1997).

4. White D.A., Jones C. \& Forman W. An investigation of cooling flows
and general cluster properties from an X-ray image deprojection analysis
of 207 clusters of galaxies. \MN\ {\bf 292}, 419--467 (1997).

5. David, L., Arnaud, K.A., Forman, W., \& Jones, C. Einstein
observations of the Hydra-A cluster and the efficiency of galaxy
formation in groups and clusters. \ApJ\ {\bf 356}, 32--40 (1990).

6. Evrard, A.E. \& Henry, J.P. Expectations for X-ray cluster observations
by the ROSAT satellite. \ApJ\ {\bf 383}, 95--103 (1991).

7. Jones, C. \& Forman, W. The structure of clusters of galaxies observed
with Einstein. \ApJ\ {\bf 276}, 38--55 (1984).

8. David, L.P., Jones, C., \& Forman, W. ROSAT PSPC observations of cool
rich clusters.
\ApJ\ {\bf 473}, 692--706 (1996).

9. Knight P.A. \& Ponman T.J. The Properties of the Hot Gas in Galaxy Groups
and Clusters from 1-D Hydrodynamical Simulations -- I. Cosmological Infall
Models. \MN\ {\bf 289}, 955--972 (1997).

10. Evrard, A.E. The intracluster gas fraction in X-ray clusters:
constraints on the clustered mass density. \MN\ {\bf 292}, 289--297 (1997).

11. Fukazawa Y. {\it et al}. ASCA measurements of silicon and iron abundances
in the intracluster medium. \PASJ\ {\bf 50}, 187--193 (1998).

12. Cavaliere A., Menci N. \& Tozzi P. The luminosity-temperature relation
for groups and clusters of galaxies. \ApJ\ {\bf 484}, L21--L24 (1997).

13. Metzler C.A. \& Evrard A.E. Simulations of galaxy clusters with and
without winds I. -- the structure of clusters. \ApJ\ (submitted)
astro-ph/9710324.

14. Watt M.P., Ponman T.J., Bertram D., Eyles C.J., Skinner G.K. \& Willmore
A.P. The morphology and dark matter distribution of the Coma cluster of
galaxies from X-ray observations. \MN\ {\bf 258}, 738--748 (1992).

15. Copi, C.J., Schramm, D.N., \& Turner, M.S. Big-Bang nucleosynthesis
and the baryon density of the Universe. \Sc\ {\bf 267}, 192--199 (1995).

16. Miralda-Escud\'e, J., Cen, R., Ostriker, J.P., \& Rauch, M. 
The Lyman-alpha forest from gravitational collpase in the cold dark
matter plus lambda model. \ApJ\
{\bf 471}, 582--616 (1996).

17. Madau, P., Pozzetti, L., Dickinson, M. The Star Formation History of
Field Galaxies. \ApJ\ {\bf 498}, 106 (1998).

18. Heckman, T.M., Armus, L.M., Miley, G.K. 
On the nature and implications of starburst-driven galactic superwinds.
\ApJS\ {\bf 74} 833--868 (1990).

19. Bower R.G. Entropy-driven X-ray evolution of galaxy clusters. \MN\ {\bf
288}, 355--364 (1997).

20. Mushotzky, R.F. \& Scharf, C.A.
The Luminosity-Temperature relation at z=0.4 for clusters of galaxies.
\ApJ\ {\bf 482}, L13--16 (1997).

21. Arnaud M. \& Evrard A.E. The $L_X-T$ relation and intracluster gas
fractions of X-ray clusters. \MN\ (submitted) astro-ph/9806353

22. Cen R. \& Ostriker J.P. Most of the ordinary matter in the Universe
is in warm/hot gas.  \Sc\ (submitted) astro-ph/9806281

23. Snowden S.L., McCammon D., Burrows D.N. \& Mendenhall J.A. Analysis
procedures for ROSAT XRT PSPC observations of extended objects and diffuse
background. \ApJ\ {\bf 424}, 714--728 (1994).

24. Ponman T.J., Bourner P.D.J., Ebeling H. \& B\"ohringer H. A ROSAT survey
of Hickson's compact galaxy groups. \MN\ {\bf 283}, 690--708 (1996).

25. Yamashita K. in {\it Frontiers of X-ray Astronomy} (eds Tanaka Y. \&
Koyama K.) 475--480 (Universal Academy Press, Tokyo, 1992).

26. Butcher J.A., PhD thesis, University of Leicester (1994).

27. Mulchaey J.S., Davis D.S., Mushotzky R.F. \& Burstein D. The intra-group
medium in poor groups of galaxies. \ApJ\ {\bf 456} 80--97 (1996).

28. Ebeling H., Mendes de Oliveira C. \& White D.A. A2572 and HCG94 --
galaxy clusters but not as we know them: an X-ray case study of optical
misclassifications. \MN\ {\bf 277}, 1006--1032 (1995).

29. Raymond J.C. \& Smith B.W. Soft X-ray spectrum of a hot plasma.
\ApJS\ {\bf 35}, 419--439 (1977).

30. Eke V.R., Navarro J.F.\& Frenk C.S. The Evolution of X-ray Clusters in
Low Density Universes. \ApJ\ (in press)  astro-ph/9708070.

{\bf Acknowledgements}\\[5mm] We thank Richard Bower, Simon White and
Peter Willmore for helpful discussions.  Data analysis was performed
on the Starlink node at Birmingham. JFN acknowledges the hospitality
of the Max Planck Institut f\"ur Astrophysik in Garching during the
preparation of this manuscript.\\[5mm]

Correspondence should be addressed to T.J.P. (email:
tjp@star.sr.bham.ac.uk).\\[5mm]

{\bf Figure Captions} 

{\bf Fig.1} --- Scaled X-ray surface brightness profiles are overlaid to show
departures from similarity in galaxy systems of different temperatures. ROSAT
PSPC images in the 0.4-2\keV\ band of the 25 systems (A262, A478, A496, A548S,
A665, A1060, A1413, A1651, A1689, A1795, A1837, A2142, A2163, A2199, A2218,
AWM4, AWM7, MKW3, MKW4, HCG62, HCG94, HCG97, NGC533, NGC2300, NGC4261) were
extracted from the public archive, corrected$^{23}$ for the effects of
vignetting and obscuration by detector support structures, and background flux
subtracted. Point sources were removed, and X-ray surface brightness profiles
constructed relative to the centroid of the remaining diffuse emission (making
due correction for areas removed). Mean temperatures for each system are taken
from GINGA and ROSAT observations$^{24-28}$. These are used to compute virial
radii$^{3}$, $r_v=0.57 (T/\keV)^{1/2} \, h^{-1}$~Mpc. The X-ray profiles have
been converted from ROSAT count rates to bolometric X-ray intensities using a
hot plasma code$^{29}$, and have been scaled by $T^{1/2} (1+z)^{9/2} \Lambda(T)$
to allow for self-similar scaling with mass and cosmic density evolution
[$\Lambda(T)$ is the temperature dependence of the hot plasma emissivity], and
by $(1+z)^{-4}$ to correct for cosmological dimming. The profiles are
colour-coded by system temperature. Similarity appears to hold for hot ($T>4$
keV) systems, but large systematic deviations are noticeable in cooler systems.

{\bf Fig.2} --- The gas ``entropy'' (defined as $S=T/n_e^{2/3}$, where T
is the mean temperature and $n_e$ is the electron number density) at a
fiducial radius $r=0.1 \, r_v$ is shown as a function of temperature for
the 25 systems in our sample. Entropies are computed by fitting
isothermal-$\beta$ models$^{7}$ to the surface brightness profiles shown
in Figure 1. Error bars indicate $90\%$ confidence levels in temperature,
and span the variation in entropy from $0.05$ to $0.2 \, r_v$. The solid
line shows the similarity relation obtained from numerical simulations
including non-radiative gas dynamics$^{30}$, $S(0.1\,r_v) \sim 45 \,
(T/{\rm keV})\, (f_{gas}/0.06)^{-2/3} \, h^{-4/3}$ keV cm$^2$, where
Hubble's constant is $H_0=100 \, h$ km s$^{-1}$ Mpc$^{-1}$. The numerical
results depend only weakly on the assumed cosmological model, and provide
a very good match to the central entropy of hot ($T>4$ keV) systems for
$f_{gas} \approx 0.06 \, h^{-3/2}$. However, poor clusters and groups have
apparently been heated to higher entropy than achievable through
gravitational collapse.

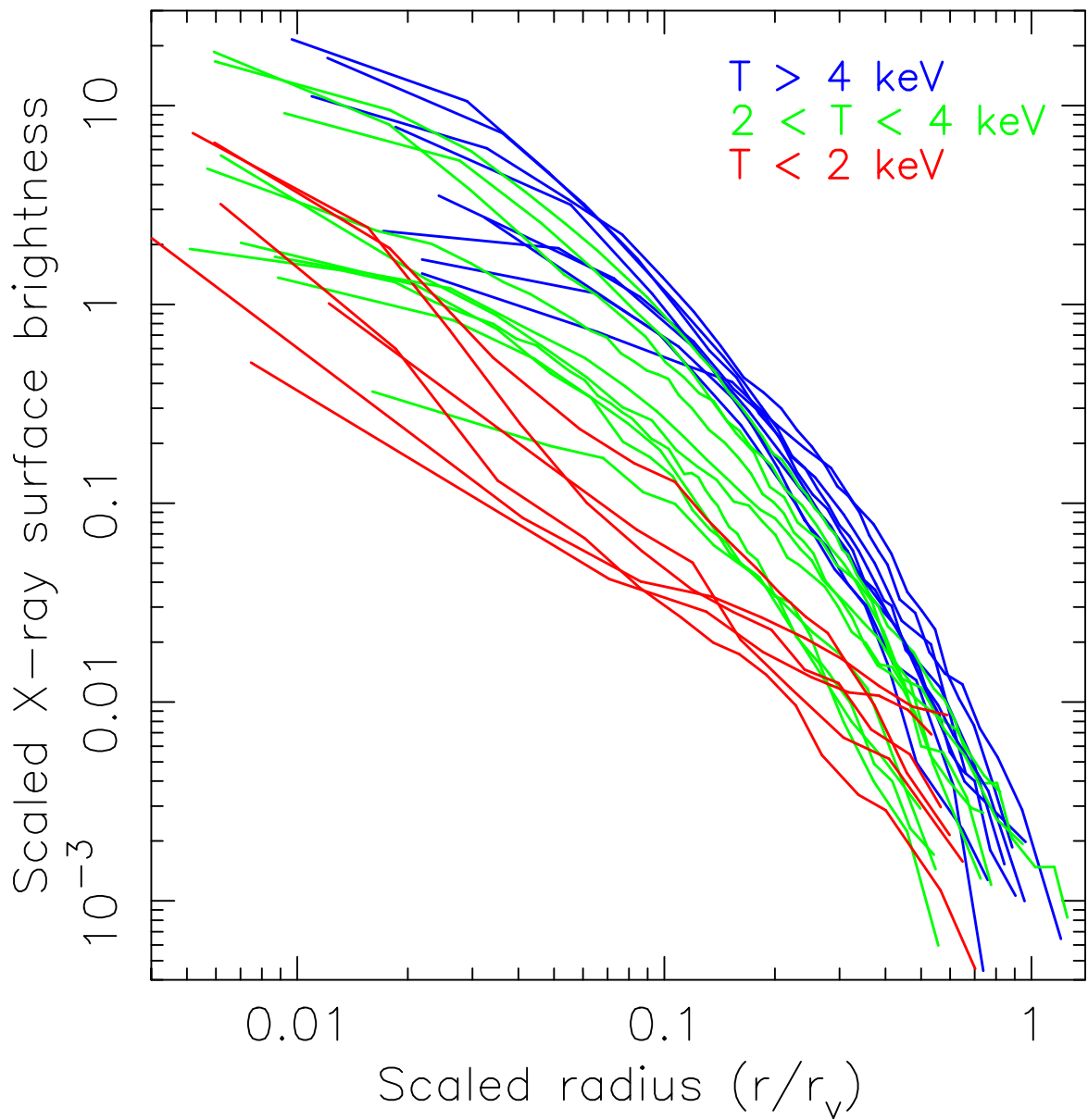
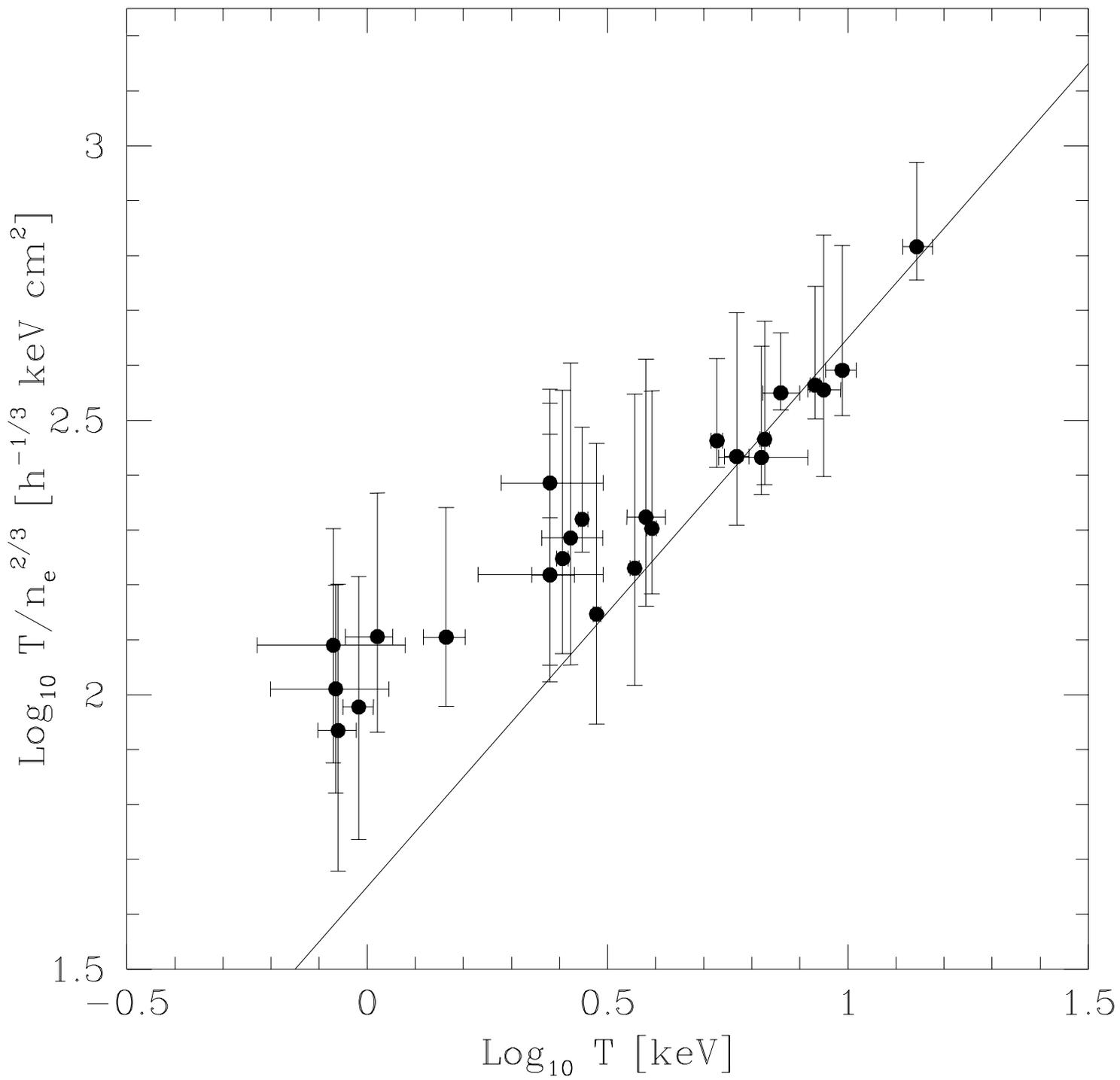
\end{document}

%% file: def.tex
\def\h50{\hbox{$h_{50}$\,}}
\def\spose#1{\hbox to 0pt{#1\hss}}
\def\ltsimm{\mathrel{\spose{\lower 3pt\hbox{$\sim$}}
	\raise 2.0pt\hbox{$<$}}}
\def\ltsim{$\mathrel{\spose{\lower 3pt\hbox{$\sim$}}
	\raise 2.0pt\hbox{$<$}}$}
\def\gtsimm{\mathrel{\spose{\lower 3pt\hbox{$\sim$}}
	\raise 2.0pt\hbox{$>$}}}
\def\gtsim{$\mathrel{\spose{\lower 3pt\hbox{$\sim$}}
	\raise 2.0pt\hbox{$>$}}$}

\def\fract#1/#2{\leavevmode\kern.1em                   
   \raise.5ex\hbox{\the\scriptfont0 #1}\kern-.1em
   /\kern-.15em\lower.25ex\hbox{\the\scriptfont0 #2}}
\def\cm{{\rm\thinspace cm}}
\def\erg{{\rm\thinspace erg}}

\def\keV{{\rm\thinspace keV}}

\def\Lsol{\hbox{$\thinspace L_{\odot}$}}

\def\Msol{\hbox{$\thinspace M_{\odot}$}}
\def\pc{{\rm\thinspace pc}}
\def\s{{\rm\thinspace s}}

\def\pcm2{\hbox{$\cm^{-2}\,$}}
\def\ergpcm3ps{\hbox{$\erg\cm^{-3}\s^{-1}\,$}}

\def\Lsolppc3{\hbox{$\Lsol\pc^{-3}\,$}}
\def\Msolppc3{\hbox{$\Msol\pc^{-3}\,$}}